\documentclass[twocolumn,prl,floatfix]{revtex4}
\usepackage{psfig}

\newcommand{\C}{{\cal C}}
\newcommand{\av}{{\mathrm{av}}}

\begin{document}
\title{Lack of Ultrametricity in the Low-Temperature phase of
3D Ising Spin Glasses}
\author{Guy Hed}
\affiliation{Department of Materials and Interfaces,
Weizmann Institute of Science, Rehovot 76100, Israel}
\author{A. P. Young}
\affiliation{Department of Physics, University of California,
             Santa Cruz, CA 95064}
\author{Eytan Domany}
\affiliation{Department of Physics of Complex Systems,
Weizmann Institute of Science, Rehovot 76100, Israel}

\begin{abstract}
We study the low-temperature spin-glass phases of the
Sherrington-Kirkpatrick (SK) model and of the 3-dimensional short
range Ising spin glass (3dISG). By using clustering to focus on
the relevant parts of phase space and reduce finite size effects,
we found that for the SK model ultrametricity becomes clearer as
the system size increases, while for the short-range case our
results indicate the opposite, i.e. lack of ultrametricity.
Another method, which does not rely on clustering, indicates that
the mean field solution works for the SK model but does not apply
in detail to the 3dISG, for which stochastic stability is also violated.
\end{abstract}

\maketitle Spin glasses constitute an example of a fascinating
family of physical systems, in which the combination of disorder
and frustration produces a multitude of low-lying states,
unrelated by symmetry. A major breakthrough came with Parisi's
mean-field solution \cite{Parisi87} of
the infinite-range Sherrington-Kirkpatrick (SK) model \cite{SK75};
a complex low-temperature phase with rich structure was revealed.
One of the central open questions of the field is whether the
qualitative structure of the mean-field solution is valid also for
the experimentally relevant 3-dimensional short-range Ising Spin
Glasses (3dISG). Both systems are described by the $N$-spin
Hamiltonian
\begin{equation}\label{eq:H}
H(\vec s) = -\sum_{\langle i,j \rangle} J_{ij} s_i s_j \;,
\end{equation}
where $s_i = \pm 1$.
In the SK model the sum in Eq.~(\ref{eq:H}) runs over all
pairs of spins, while in the 3dISG the sum is over nearest neighbor
sites of a cubic lattice. The interactions $J_{ij}$ are taken from
a Gaussian distribution with mean zero and variance $1/(N-1)$ in the SK
model, while the variance is 1 for the 3dISG.

To test the validity of the mean field solution, it is natural to
consider first the most widely calculated observable in ISG; the
distribution $P(q)$ of the overlaps $q$, defined for two states
$\vec s^\mu$ and $\vec s^\nu$ as $q_{\mu\nu}= N^{-1} \sum_{i=1}^N
s^\mu_i s^\nu_i$. For each realization of the randomly selected
$J_{ij}$ one calculates the probability distribution $P_J(q)$ (at
equilibrium at some temperature); $P(q)=[P_J(q)]_\av$ is
the average over the different realizations $\{ J \}$ of the
disorder. Indeed, $P(q)$ was measured by Monte-Carlo simulations
\cite{Marinari98} and in other ways \cite{Krzakala00,Palassini00}
and non-trivial overlap distributions were found for both the SK
and 3dISG. However, there were clear indications
that the low $T$ phases of the two models were
different \cite{katzgraber01,SCHICS}. In particular, evidence was
presented \cite{SCHICS} for lack of ultrametricity in 3dISG,
and for the central role played by correlated macroscopic
spin domains in producing a non-trivial $P(q)$.

Ultrametricity (UM) of the state space \cite{Mezard84} is one of
the main characteristics of the low $T$ phase of the mean-field
solution and the SK model. Consider an equilibrium ensemble of
microstates at $T<T_c$, and pick three, $\rho$, $\mu$ and $\nu$ at
random (see, however, point (a) below). These indices represent
the states $\vec s^\rho$, $\vec s^\mu$ and $\vec s^\nu$. Order
them so that $q_{\mu\nu}\geq q_{\nu\rho}\geq q_{\mu\rho}$;
ultrametricity means that in the thermodynamic limit we get
$q_{\nu\rho}=q_{\mu\rho}$ with probability 1. In terms of the
distances $d_{\mu \nu}=(1-q_{\mu \nu})/2$ the condition of UM
becomes $d_{\nu\rho}=d_{\mu\rho} \geq d_{\mu\nu}$.

Even though claims were made for the emergence of UM for large 3D
systems \cite{Franz00}, these were not conclusive. Since
demonstrating lack of ultrametricity suffices to prove that the
mean-field solution does not apply for the 3dISG, the evidence
presented in \cite{SCHICS} prompted Ciliberti and Marinari (CM)
\cite{Ciliberti03} to use similar methods at the same system sizes
as used by \cite{SCHICS}, to search for ultrametricity in the SK
model, which has an ultrametric low-T phase. CM did not see UM in
the SK model and claimed that the system sizes used were too small
to allow observation of ultrametricity. CM argued further that
failure to see UM where it holds indicates that similar failure
reported \cite{SCHICS} for the 3dISG does not provide evidence for
its absence.

\begin{figure}[t]
\centerline{\psfig{figure=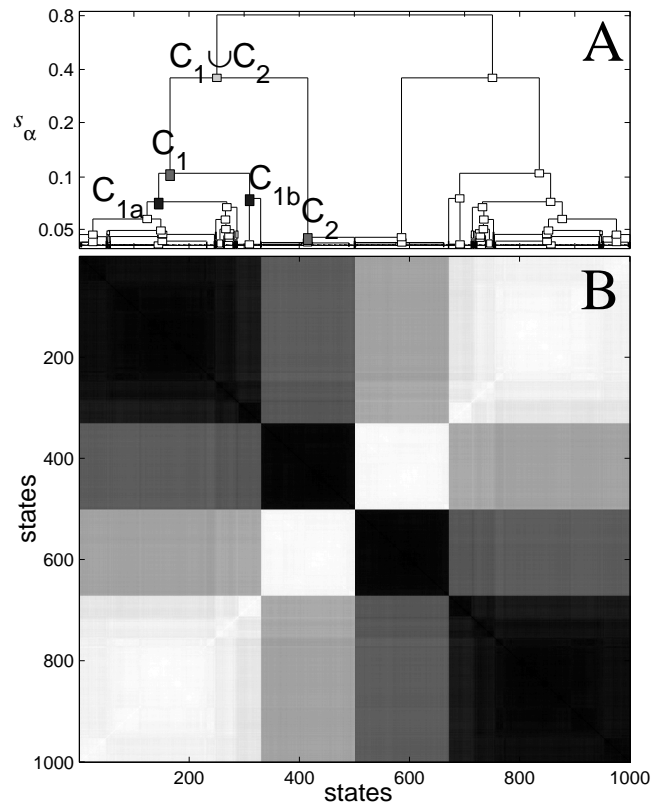,width=\columnwidth}}
\caption{(A)
Clustering produces the dendrogram of state clusters, for the
equilibrium ensemble of 1000 states, obtained by simulation of one
realization of the SK model for 1024 spins, at $T=0.2$. The
vertical position of each cluster $\alpha$ is its score
$s_\alpha$. Logarithmic scale is used for the vertical axis; only
clusters with $s_\alpha > 0.04$ are shown. Significant clusters,
identified by a long branch above them, have high values of
$v_\alpha$; these clusters reflect presence of dense regions in
the underlying Boltzmann distribution. (B) The distance matrix
$d_{\mu\nu}$ of the microstates, that were reordered according to
their positions, as the leaves of the dendrogram, on the
horizontal axis of (A). Darker colors indicate smaller distances.}
\label{figDEND}
\end{figure}

In order to settle this issue we performed extensive simulations
of the SK model for a somewhat larger range of sizes than in CM, and carried
out a
careful analysis of the results. We also applied
the same method of analysis to our
earlier~\cite{SCHICS} results for the 3dISG model.
As opposed
to CM, we do find clear evidence for UM in the SK model; it's
signature becomes more pronounced as the system size increases. On
the other hand, for the 3dISG we find the opposite; with
increasing system size the results become less consistent with UM.
We will explain in detail elsewhere \cite{future} the reasons for
the differences between our results and those of CM. Here we note
that in order to observe UM one has to avoid three main pitfalls:
\begin{itemize}
\item[(a)] If time reversal symmetry is unbroken (zero field), phase
space structure consists of two spin-flip related pure-state
hierarchies (see Fig. \ref{figDEND}). UM can be observed only if
all three states of each sampled triplet  belong to the same side
\cite{Ciliberti03,MezardPC}.
\item[(b)] Do not work too close to $T_c$ otherwise finite size effects
mask the signal.
\item[(c)] At low $T$, most of the microstates belong to the same pure state
\cite{Mezard84}. With increasing $N$ the overlaps inside a pure
state converge to $q_\mathrm{EA}$ and most $\{\rho,\mu,\nu\}$
triangles become equilateral, conveying no information on UM.
\end{itemize}
To avoid
pitfalls (a) and (c) we map out the
upper levels of the pure state hierarchy by clustering
\cite{Hed01,SCHICS} (see Fig \ref{figDEND}). To avoid (a), we
consider only triplets of states from (say) the left tree and to
avoid (c) we pick the three states from different
pure states of this tree.

{\it Monte Carlo simulations:} SK systems were simulated at sizes
$N=32,128,256,1024$.
To speed up equilibration we used
parallel tempering
with 21 temperatures ranging from $2.0$ down to $0.2$.
Tests for equilibration were carried out as in Ref.~\onlinecite{katzgraber01}.
Details of the simulations on the 3dISG are given in Ref.~\onlinecite{SCHICS}.
For each
size we generated 500 different realizations $\{ J \}$ of the
disorder (except for $N=8^3$ - only 335 realizations). For each
$\{ J \}$ we generated 500 microstates, sampled according the
Boltzmann distribution at $T=0.2$. Each sample of states was
enhanced to 1000 by adding to these 500 microstates their
spin-flip mirror images. The result is two spin-flip related
unbiased samples - one from each tree. The microstates appear in
the random order generated by the Monte Carlo procedure.

\begin{table*}[!tbp]
\begin{center}
\begin{tabular}{|c|c|c||c|c|c||c|c|}
\hline\hline
\multicolumn{3}{|c||}{3dISG} & \multicolumn{3}{c||}{SK}
& \multicolumn{2}{c|}{Naive SK} \\
\hline
$N$ & $\overline{K}$ & $P(K=1)$ & $N$ & $\overline{K}$ & $P(K=1)$
& $\overline{K}$ & $P(K=1)$ \\
\hline\hline
$4^3$ & 0.357(12) & 0.617(16) &  32  & 0.331(9) & 0.350(16) & 0.316(6) & 0.759(7) \\
$5^3$ & 0.576(12) & 0.233(13) & 128  & 0.346(8) & 0.037(5)  & 0.405(4) & 0.365(7)  \\
$6^3$ & 0.548(10) & 0.144(9)  & 256  & 0.294(7) & 0.005(1)  & 0.423(3) & 0.127(4) \\
$8^3$ & 0.502(12) & 0.044(5)  & 512  & 0.236(6) & 5(3) $\times 10^{-4}$ & 0.395(2) & 0.015(1)\\
    &             &           & 1024 & 0.182(5) & 2(1) $\times 10^{-6}$ & 0.345(3) & 1.8(2) $\times 10^{-4}$ \\
\hline\hline
\end{tabular}
\end{center}

\caption{The values of $\overline K$ as a function of $N$, for
both the SK and 3dISG. See text for definitions.}
\label{tabK}
\end{table*}

{\it Identification of pure states by clustering:} The 1000
microstates, obtained for a realization of the disorder $\{ J \} $,
are assigned to groups with a relative tree-like
structure \cite{SCHICS} by clustering. Here~\cite{note2}
we used the
well known {\it average linkage} agglomerative clustering algorithm
\cite{Jain88}. Initially each of the $M=1000$ points (microstates)
is a cluster; at every clustering step those two clusters $\alpha,
\beta$ that are closest are merged to form a new cluster $\gamma$;
the process stops after $M-1$ steps when all the points belong to
one cluster. The distance $d_{\alpha, \beta}$ between clusters is
defined as the average distance between points of $\alpha$ and
$\beta$. After every merging operation we record for the new
cluster $\gamma=\alpha\cup\beta$ the score $s_\gamma
= d_{\alpha, \beta}$, and for each of the merged clusters a
quality measure $v_\alpha=s_\gamma/s_\alpha $. If $v_\alpha$ is
high, the distances between states within $\alpha$ are
significantly smaller then the distances of these states from
states outside $\alpha$, and cluster $\alpha$ is ``real", i.e. not
an artifact of the agglomerative procedure.

Fig. \ref{figDEND} presents the tree of clusters (dendrogram)
obtained when the procedure outlined above was applied to the
equilibrium ensemble of 1000 microstates, obtained at $T=0.2$ for
a particular realization $\{ J \}$ of the SK model with $N=1024$ spins.
The clusters at the higher levels of
the hierarchy have high $v_\alpha$ values, suggesting that the
partitions they form indeed exist in the underlying distribution
from which the states were taken. That is, they map the
microstates in our sample to the pure states from which they were
taken.

At the highest level, the dendrogram splits into two identical
trees, which are the two spin-flip related pure state trees. We
consider the states that belong to one (the left). For all $\{ J
\}$ we call  the  two top level state clusters by $\C_1$ and
$\C_2$ with the convention $|\C_1|\geq|\C_2|$. The two subclusters
of $\C_1$ are denoted by $\C_{1a}$ and $\C_{1b}$. The triplets of
microstates for which we test UM are selected as follows:
$\mu\in\C_{1a}$, $\nu\in\C_{1b}$ and $\rho\in\C_2$. If state space
is ultrametric and these clusters correspond to the underlying
pure states, we expect $d_{\mu\rho}=d_{\nu\rho} \geq d_{\mu\nu}$ .

\begin{figure}[b]
\psfig{figure=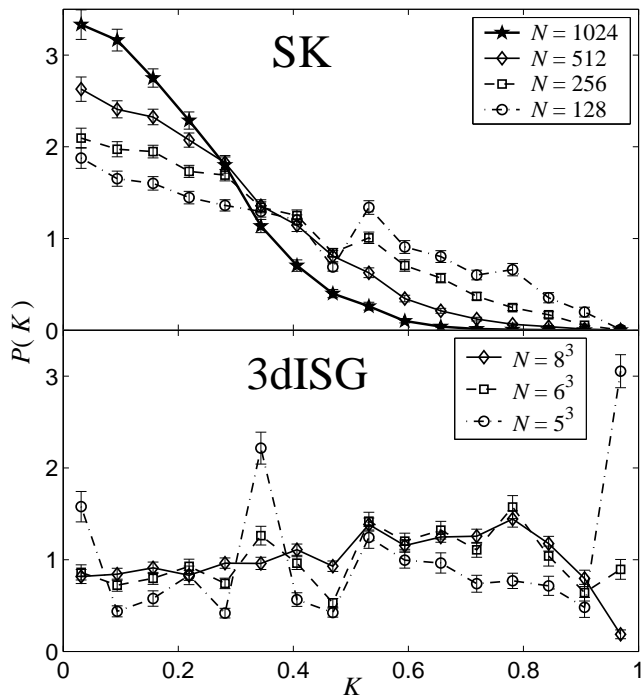,width=\columnwidth}
\caption{The distributions
$P(K)$ (see text), calculated for 3dISG and SK systems for various
sizes $N$.}
\label{figK}
\end{figure}

{\it Direct Measurement of Ultrametricity:} For each triplet of
states $\mu$, $\nu$ and $\rho$ selected as described above, we
define an index \cite{SCHICS}
\begin{equation}\label{eq:K}
K_{\mu\nu\rho} = \frac{ |d_{\mu\rho}- d_{\nu\rho}| }{ d_{\mu\nu} }
\;.
\end{equation}
Note that $0\leq K_{\mu\nu\rho}\leq 1$ due to the triangle inequality.

For a given realization $\{ J \}$ we identify $\C_{1a}$,
$\C_{1b}$, and $\C_2$, and measure $K_{\mu\nu\rho}$ over all the
corresponding state triplets. We discard triplets with $K=1$ since
these correspond to co-linear ``triangles" -- they are the result
of a finite size effect, and their weight $P(K=1)$ decreases
rapidly with increasing system size, as seen in Table \ref{tabK}.
We calculate the distribution $P_J(K)=P_J(K_{\rho\mu\nu} =
K|K_{\rho\mu\nu}<1)$, and its mean $K_J$. We then average over
the realizations $\{ J \}$ to get $\overline{K}=[K_J]_\av$
and $P(K) = [P_J(K)]_\av$. If state space is UM in the
thermodynamic limit, we expect $\overline K \rightarrow 0 $ and
$P(K) \rightarrow \delta(K)$ as $N \rightarrow \infty$.

We present in Fig. \ref{figK} the distributions $P(K)$ obtained
for a sequence of sizes $N$ for the SK and 3dISG models.  For the
SK model the peak of $P(K)$ is at $K=0$ and it's width decreases
as $N$ increases, as expected if $P(K)$ is to approach
$\delta(K)$. The behavior is markedly different for the 3dISG: the
width of the distribution remains constant and its peak is at
$K\simeq 0.8$. For the SK model the values of $\overline K$,
presented in Table \ref{tabK}, are smaller, decrease faster and
approach zero as $N$ increases, whereas for the 3dISG $\overline
K$ seems to go to a non-vanishing limiting value. For both systems
the lowest size is useless for estimating the trend with
increasing $N$.

We present also results obtained for the SK model by a ``naive"
calculation, that neglects to ensure that all members of a triplet
of states are drawn from the same side of the
dendrogram;
such oversight indeed renders UM unobservable, as also noted by
\cite{Ciliberti03,MezardPC}.

\begin{table}[b]
\begin{center}
\begin{tabular}{|c|c|c||c|c|c|}
\hline\hline
\multicolumn{3}{|c||}{3dISG} & \multicolumn{3}{c|}{SK} \\
\hline
$N$ & $[s_{\C_1\C_2}]_\av$ & $[S_{\C_1\C_2}]_\av$ &
$N$ & $[s_{\C_1\C_2}]_\av$ & $[S_{\C_1\C_2}]_\av$ \\
\hline\hline
$4^3$ & 0.050(2) & 0.135(7) &  32  & 0.098(3) & 0.338(29) \\
$5^3$ & 0.045(2) & 0.123(6) & 128  & 0.064(2) & 0.196(12) \\
$6^3$ & 0.043(2) & 0.133(10)& 256  & 0.054(2) & 0.170(10) \\
$8^3$ & 0.042(2) & 0.136(9) & 512  & 0.043(1) & 0.142(10) \\
      &          &          & 1024 & 0.034(1) & 0.115(10) \\
\hline\hline
\end{tabular}
\end{center}
\caption{The standard deviation $s_{\C_1\C_2}$ and the normalized
standard deviation $S_{\C_1\C_2}$
of the elements of $q_{\C_1\C_2}$ (see text), averaged over
realizations of the disorder $\{ J \}$, for 3dISG and SK systems
of different sizes.} \label{tabC12}
\end{table}

Another test of UM is the following: if the state space is
ultrametric, then the distance between states $\mu\in\C_1$ and
$\nu\in\C_2$ depends only on the identity of their closest common
ancestor on the  tree of pure states, which is $\C_1\cup\C_2$.
Thus, for each realization, the sub-matrix $q_{\C_1\C_2}$ should
be uniform (e.g. the rectangle in the distance matrix of Fig.
\ref{figDEND}, that connects states from $C_1$ and $C_2$, should
be uniformly shaded).

We have calculated the standard deviation $s_{\C_1\C_2}$ and
the average $m_{\C_1\C_2}$of the
elements of $q_{\C_1\C_2}$ for each realization $\{ J \}$.
For a given average $m$ the standard deviation is bounded by
$\sqrt{m(1-m)}$ - the standard deviation of a Bernoulli distribution.
We define a normalized standard deviation
$S_{\C_1\C_2}=s_{\C_1\C_2}/\sqrt{m_{\C_1\C_2}(1-m_{\C_1\C_2})}$.
We average over the realizations to get $[s_{\C_1\C_2}]_\av$
and $[S_{\C_1\C_2}]_\av$.
The results are given in Table \ref{tabC12}. For the SK model
both $[s_{\C_1\C_2}]_\av$ and $[S_{\C_1\C_2}]_\av$
decrease rapidly as $N$ increases, indicating again that UM is approached,
while for the 3dISG the values of $[s_{\C_1\C_2}]_\av$ and
$[S_{\C_1\C_2}]_\av$ seem to asymptote to 0.04 and 0.13
respectively.

\begin{figure}[b]
\psfig{figure=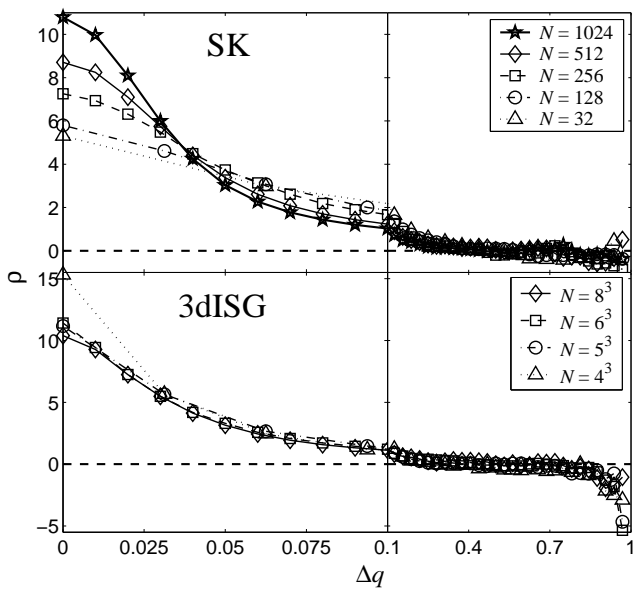,width=\columnwidth}
\caption{The function $\rho(\Delta q)$ for SK and 3dISG systems.
The error bars are smaller than the symbols.}
\label{fig3}
\end{figure}

{\it To test the validity of the mean field solution without using
clustering methods}, we focused on one of several results for
multi overlap distributions. M{\'e}zard et al. \cite{Mezard84}
showed that the distributions $P_J(q)$ of ISG satisfy the relation
\begin{equation}
\label{eq:pp}
[P_J(q_1)P_J(q_2)]_\av = \frac{2}{3}P(q_1)P(q_2) +
\frac{1}{3}P(q_1)\delta(q_1-q_2) \;.
\end{equation}
which, just like UM, is a consequence of Parisi's mean field
solution. Eq. (\ref{eq:pp}) has been derived for the SK model
without replicas by Guerra~\cite{Guerra96} and, more generally, is
a consequence of ``stochastic
stability''~\cite{Aizenman98,Parisi00}. In order to test the
validity of Eq.~(\ref{eq:pp}) we define
\begin{equation}\label{eq:rho}
\rho(q_1,q_2) = \frac{ 3[P_J(q_1)P_J(q_2)]_\av - 2P(q_1)P(q_2) }{P(q_1)} \;.
\end{equation}
We study $\rho(\Delta q)$ - the average of $\rho$ over overlap
pairs $\{q_1,q_2\}$ with $|q_1-q_2| = \Delta q$. As $N\to\infty$
we expect $\rho$ to have a $\delta$-peak at $\Delta q =0$
\cite{ParisiPC}. In general, this peak can have a small
weight and $\rho$ may have 
a continuous part, while in the mean
field solution $\rho(\Delta q)$ converges
to $\delta(\Delta q)$.

Fig. \ref{fig3} presents $\rho(\Delta q)$. For the SK model $\rho$
approaches $\delta (\Delta q)$ with increasing $N$, while for
the 3dISG the height of the peak at 0 decreases and seems to
converge to a finite value, and the width remains constant. For
this case we conclude that $\rho$ is the sum of a $\delta(\Delta
q)$-peak, and a continuous term; the rise of the first with $N$ is
masked by the decrease of the second.

We also examined the equality of the second moment of Eq.~(\ref{eq:pp}):
$[\langle q^2\rangle^2]_\av = \frac{2}{3}{[\langle q^2\rangle]_\av}^2
+ \frac{1}{3}[\langle q^4\rangle]_\av$, which has been investigated
numerically in other work~\cite{Marinari00}. We find that the ratio of the
two sides is one for 
both SK and 3dISG of all sizes
within an error smaller than 0.03, in agreement with
Ref.~\onlinecite{Marinari00}. However, moments may be insensitive to
deviations from Eq.~(\ref{eq:pp}) because
$\rho\approx 0$ when $|q_1-q_2|>0.2$.

Top conclude, we have
presented strong evidence for the lack of UM in the low $T$
phase of the 3-dimensional short range Ising spin glass.
We conclude
that our methods are able to detect UM, where it
exists, at the system sizes studied. Our findings indicate that
the structure of the low-temperature phase of spin glasses
that emerges from the mean-field solution is not valid for the
3-dimensional short range Ising spin glass.

\begin{acknowledgments}
We thank  E.~Marinari and I.~Kanter for helpful correspondence and
discussions and particularly, G. Parisi for generously sharing his
insights and correcting inaccuracies of our preprint. This work
was supported by grants from the Minerva Foundation,
the Germany-Israel Science Foundation
and the European Community's Human Potential Programme under contract
HPRN-CT-2002-00319, STIPCO. The work of APY was supported by NSF
grants DMR 0086287 and 0337049.
\end{acknowledgments}

\newpage
\printtables
\newpage
\printfigures


\begin{thebibliography}{20}
\bibitem[{not()}]{note2}
\bibinfo{note}{In Ref.~\onlinecite{SCHICS} we used Ward's clustering method,
  which tends to ignore pure states with small statistical weight. The present
  scheme is more suitable for our data, but the results of the analysis are
  similar for both schemes.}
\end{thebibliography}
\end{document}